%% file: main.tex
\pdfoutput=1

\documentclass[11pt]{article}
\usepackage{booktabs}

\usepackage[preprint]{acl}
\usepackage{ragged2e}
\usepackage{times}
\usepackage{latexsym}
\usepackage{amsmath}
\usepackage{appendix}
\usepackage{tabularx}
\usepackage{multicol}
\usepackage{multirow}
\usepackage{listings}
\definecolor{codegreen}{rgb}{0,0.6,0}
\definecolor{codegray}{rgb}{0.5,0.5,0.5}
\definecolor{codepurple}{rgb}{0.58,0,0.82}
\definecolor{backcolour}{rgb}{0.95,0.95,0.95}

\lstdefinestyle{mystyle}{
    backgroundcolor=\color{backcolour},
    commentstyle=\color{codegreen},
    keywordstyle=\color{magenta},
    numberstyle=\tiny\color{codegray},
    stringstyle=\color{codepurple},
    basicstyle=\ttfamily\footnotesize,
    breakatwhitespace=false,         
    breaklines=true,                 
    captionpos=b,                    
    keepspaces=true,                 
    numbers=left,                    
    numbersep=5pt,                  
    showspaces=false,                
    showstringspaces=false,
    showtabs=false,                  
    tabsize=2
}

\usepackage[T1]{fontenc}

\usepackage[utf8]{inputenc}

\usepackage{microtype}

\usepackage{inconsolata}

\usepackage{graphicx}

\newcommand{\datasetname}{AmazonQAC}
\newcommand{\upperbound}{69.8\%}

\title{\datasetname: A Large-Scale, Naturalistic Query Autocomplete Dataset}

\author{
\textbf{Dante Everaert} \\
Amazon Search \\
\texttt{danteev@amazon.com} \\
\texttt{dante.everaert@gmail.com} \\
\And
\textbf{Rohit Patki} \\
Amazon Search \\
\texttt{patkir@amazon.com} \\
\And
\textbf{Tianqi Zheng} \\
Amazon Search \\
\texttt{tqzheng@amazon.com} \\
\AND
\textbf{Christopher Potts} \\
Stanford University \\
\texttt{cgpotts@stanford.edu}
}

\begin{document}
\maketitle
\begin{abstract}
Query Autocomplete (QAC) is a critical feature in modern search engines, facilitating user interaction by predicting search queries based on input prefixes. Despite its widespread adoption, the absence of large-scale, realistic datasets has hindered advancements in QAC system development. This paper addresses this gap by introducing \datasetname, a new QAC dataset sourced from Amazon Search logs, comprising 395M samples. The dataset includes actual sequences of user-typed prefixes leading to final search terms, as well as session IDs and timestamps that support modeling the context-dependent aspects of QAC. We assess Prefix Trees, semantic retrieval, and Large Language Models (LLMs) with and without finetuning. We find that finetuned LLMs perform best, particularly when incorporating contextual information. However, even our best system achieves only half of what we calculate is theoretically possible on our test data, which implies QAC is a challenging  problem that is far from solved with existing systems. This contribution aims to stimulate further research on QAC systems to better serve user needs in diverse environments. We open-source this data on Hugging Face at \url{https://huggingface.co/datasets/amazon/AmazonQAC}.
\end{abstract}

\section{Introduction}

Query Autocomplete (QAC) is an important feature in nearly every modern search engine \citep{ir-qac2}. As the user types out a search query, the QAC system’s aim is to provide a list of search term suggestions based on the partially typed query (the “prefix”). Ideally, the QAC system will provide the user’s intended query, which they can select, thereby saving them the effort of typing out the full query. Even where the user does not have a specific query in mind, QAC suggestions can help them formulate search queries that lead them to the results they are seeking.

However, despite the importance of QAC, it is a comparatively under-explored task in research. The publicly available datasets tend to be derived from search query datasets \citep[e.g.][]{Patki2024, trienlg,park2017neural}. However, these datasets do not contain the prefixes that users typed, so  prefixes have to be synthetically constructed \citep{synth1}, greatly limiting the empirical value of these resources for QAC. In fact, we were unable to find any publicly available large scale QAC datasets beyond synthetically constructed ones from the AOL data release in 2006. In general, the lack of large-scale realistic benchmarks has hampered research on QAC; few tasks have as large a gap between their importance in real-world technologies and the amount of research devoted to them.

\input{table-examples.tex}

In this paper we aim to facilitate more research on QAC by releasing \datasetname, a QAC dataset collected from Amazon Search logs with participation and support from Amazon. \datasetname\ contains 395M anonymized examples, where an example consists of a submitted search term together with the sequence of prefixes that was typed to reach that search term, a session ID, a timestamp, and other metadata (Table~\ref{tab:example_table}). The session IDs and timestamps mean that multiple sequential user searches can be grouped together to form context, which has shown to be useful for QAC \citep[e.g.][]{context-qac1,context-qac2}. The dataset includes a test set of 20K examples from a later time period than the train set, designed to simulate a real-world deployment of a QAC system.

We present the task description, analyze dataset statistics, describe evaluation metrics, and motivate an upperbound Success@10 score of \upperbound\ on our held-out test set. We also evaluate several baseline approaches on \datasetname: Prefix Trees, semantic retrieval, and Large Language Models (LLMs) with and without finetuning. We find that the QAC problem is not just a simple case of prefix-search term memorization, as conventional wisdom might imply, but rather that it is a complex recommendation problem that is significantly influenced by the user's search context. Our best baseline system is a finetuned LLM that leverages session context, and it achieves Success@10 of 37, which is half of our upperbound. This indicates that the QAC problem is a difficult one not readily solved by current systems.
We hope that releasing \datasetname\ will prompt further innovation in QAC systems and that our baseline systems help guide these research efforts.

\section{Task Description and Data Preparation}

\subsection{Task Description}

Broadly, a QAC system in a search engine provides a list of relevant search term suggestions given the current user-typed prefix, as the user types their intended search. There are two core mandates a QAC system should serve: given a prefix input $p$ and the user's intended final search term $s$ (which may or may not be a string-literal completion of the prefix), the QAC system's main goal is to provide $s$ in a list of $N$ suggestions (usually 10) that the user sees in the interface, with the secondary aim of placing $s$ as high as possible in the list. In practice, we expand the definition to allow for other contextual inputs like past searches ($c$) which could be useful to predict $s$. Thus, given a set of $(c, p, s)$ tuples, the QAC task is to optimize for the presence and rank of $s$ given $(c, p)$ in the QAC system's top $N$ provided suggestions. We give examples of the data in Table~\ref{tab:example_table}, highlighting past search, prefix and completion triplets as well as cases where the user-typed prefix matches and does not match their final search term.

\subsection{Data Preparation}\label{sec:prep}

\input{table-stats.tex}

We collect \datasetname\ from Amazon Search autocomplete customer logs in the U.S., with the support and technical assistance of Amazon. All data has been scrubbed for personally identifiable information (PII) with a wide variety of regex matches to remove any patterns commonly associated with PII (see Appendix \ref{app:regex}). We further limit the dataset to contain terms which have been searched at least 4 times by at least 4 different sessions, and filter all search terms through an LLM to flag inappropriate or personal content, as an additional measure to ensure user privacy (\ref{app:llm_filter}). The full dataset is available on Hugging Face at \url{https://huggingface.co/datasets/amazon/AmazonQAC}.

\paragraph{Main Data.} Existing QAC datasets generally include only the final search term, leading researchers to construct synthetic prefixes from that search term \citep[e.g.][]{synth1,synth2}.  In contrast, we provide both the final search term and the sequence of prefixes a user typed which led to that search term. For example, if a user typed “iph” and selected “iphone” from the QAC list, the dataset would have prefix list [“i”, “ip”, “iph”] leading to the search term “iphone”. 

We believe the synthetic approach has several key disadvantages. First, synthetically constructed prefixes assume users type out the search term in a linear manner, but we find that nearly 38\% of user typing sequences are in a non-linear pattern where the previous prefix typed in the sequence is not itself a prefix of the current one (e.g. [``i'', ``ip'', ``ipo'', ``ip'', ``iph'', ``ipho'', \ldots]). Rather, it is a deletion, or word substitution, usually due to misspellings. 

Second, advanced QAC systems go beyond strict-prefix-matching and provide semantically meaningful suggestions. We find that 13\% of final search terms are not prefixed by the final typed prefix (e.g., prefix ``ipad ca'' and search term ``case for ipad''). Such patterns are not possible to capture with the synthetic construction.

Finally, providing real prefix sequences enables modeling how much of a search term users type before selecting a QAC suggestion. 

We also collect the session ID, final search timestamp, and timestamp of the first prefix typed. These metadata allow us to reconstruct search sessions and condition model predictions on session history.

In sum, \datasetname\ consists of the following columns: search term ID, user session ID, sequence of prefixes, timestamp of first typed prefix, final search term, timestamp of final search, popularity (full schema in Appendix \ref{app:schema}). This dataset is constructed from a sample of searches from 2023-09-01 to 2023-09-30. Popularity is a count of how many times that search term appears in the dataset.

\paragraph{Main Data Analysis.} We provide detailed statistics on the dataset in Table~\ref{tab:analysis}. In the top section of the table, we provide overall statistics: number of words, number of prefixes, number of unique prefixes, number of unique final search terms, and number of unique sessions. These are useful for gaining an overall picture of the dataset. In the middle section of the table, we provide detailed pattern statistics: average final search term word length, average prefix length, percentage of final search terms which match the prefix, and number of searches per session. These are useful for understanding users' typing patterns.

We derive several important insights from the dataset statistics. First, the number of unique prefixes and the number of unique search terms are low (8.9\% and 10.0\% respectively). However, the number of unique prefix/search term pairs is much higher, at 25.9\%. In other words, while there is significant repetition in the data, users don't arrive at the same suggestion in the same way. We also find that users type approximately 48\% of the final search term before selecting the search term from the suggestion list. Finally, we find that, in 13\% of searches, users selected a QAC suggestion which did not match the prefix. This motivates QAC systems which go beyond the elementary prefix-matching paradigm.

\paragraph{Test Data.} In practice, a QAC system should be able to perform well on prefixes/search terms in the future, past the date with which any historical training data was used to build the system. To that end, we sample test data from the 2 weeks after the training data, 2023-10-01 to 2023-10-14. In addition, we construct the test set to mimic the conditions a QAC service would encounter if deployed. In practice, a QAC system receives a series of asynchronous/unrelated (prefix, context) requests and is tasked with providing search term suggestions for each request. In this setup, the QAC system would not have access to the sequence of prefixes being typed out or past suggestions provided for a sequence. To that end, the test set we provide is a sample of 20,000 random single prefix/final search term pairs from the test set along with an array of the past searches in the session for each prefix/search term pair.

\paragraph{Test Data Analysis.} We compute the same statistics on the test dataset as we do on the training dataset, shown in Table~\ref{tab:analysis}. In order to compare the test and train dataset, we additionally compute the overlap in unique prefix/search term pairs with the training dataset. This analysis is summarized in the bottom section of the table. It shows a high overlap in prefixes (88\%) and search terms (74\%), but this drops to 59\% overlap when considering prefix/search term pairs. This means that, while a QAC system trained on the main data may have seen 74--88\% of the prefixes and final search terms before, it has only seen about half of the exact prefix/search term combinations before. In terms of the search pattern changes between the main data and test data, we find a statistically significant difference (t-test, $p < 0.05$) in the average search term length, number of searches per session, and the percentage of final prefixes which match the final search term (76\% test vs 87\% train). These attribute changes confirm our hypothesis that user interaction patterns with QAC vary over time. In all, our test set's unseen prefix/completion terms and shift in statistics provide a realistic test of a QAC system’s adaptability to new scenarios.

\section{Evaluation Metrics}

\subsection{Core Metrics}

The QAC’s system has a dual mandate to provide the correct final search term in a short list and rank that search term highly in that list. This motives two metrics. For the first mandate, we use the metric of Success@10. Formally given a $(c, p, s)$ triplet and a $\textit{QAC}$ function to produce search term suggestions using $(c, p)$, Success@10($c, p, s$) is:

\begin{equation*}
    \text{Success@10}(c, p, s) =
    \begin{cases}
  1 & \text{if } s \in \textit{QAC}(c, p)@10 \\
  0 & \text{otherwise}
\end{cases}
\end{equation*}
where $\textit{QAC}(c, p)@10$ is the set of top 10 items returned by $\textit{QAC}(c, p)$. We report the average of this value over the full evaluation set.

For the ranking mandate, we use Reciprocal Rank \citep{mrr}:
\begin{equation*}
\text{RR}@10(c, p, s) =
    \begin{cases}
  \frac{1}{\text{pos}(s)} & \text{if } s \in \textit{QAC}(c, p)@10 \\
  0  & \text{otherwise}
\end{cases}
\end{equation*}
where $\text{pos}(s)$ is the position of $s$ in the ranking determined by $\textit{QAC}(c, p)$.
We again report the mean of this value over the full evaluation set (MRR@10).

\subsection{Performance Upperbound}

QAC is a difficult and often ambiguous task, as a given prefix might be compatible with numerous reasonable search terms. Thus, to help contextualize our baseline performance numbers, we now estimate an upperbound for performance on \datasetname. To do this, we make two assumptions. 

Assumption~1 is that any past search context beyond 1 hour does not provide any information for the next search. If a prefix does not have a past search context, it is not possible to disambiguate different search terms for the same prefix. (For example, we cannot systematically provide better search terms for one user prefix ``i'' over another user prefix ``i'' if neither have context). The best theoretical performance any system could do on those test set prefixes would be to provide the top 10 most popular search terms based on true observed retrospective popularity during the test set dates. 43.2\% (8,641) of our test-set search terms fall into this no-context group. For them, Success@10 is 30.1\% (2,598 successes) using true observed popularity, according to our maximally optimistic criterion.

Assumption~2 is that any past search within 1 hour provides perfect information for the next search. 56.8\% (11,359) of our test-set search terms are in this group. We assume that the best systems would be able to provide perfect suggestions for this group (100\% Success@10).

Putting the above two estimates together, we conclude that the best system would achieve an average Success@10 of \upperbound\ on our test set.

\section{Baseline Systems and Results}
In order to provide researchers with QAC baselines on our dataset, we train and benchmark a cross-section of different QAC approaches with our dataset. 
Our implemented systems may not be state-of-the-art, since we rather aim to provide a base number and insights into how different approaches to QAC behave on our dataset.

QAC approaches can broadly be split into two categories: information-retrieval (IR) QAC and generative QAC. 
We explore representative models from both categories.

Our results are summarized in Table~\ref{tab:results}.


\begin{table}[tp]
\resizebox{1\linewidth}{!}{
\setlength{\tabcolsep}{2pt}
  \begin{tabular}{@{} l c c @{}}
  \toprule
  \textbf{System} & \textbf{Success@10} & \textbf{MRR@10} \\
  \midrule
  \textbf{IR: Prefix Tree} & 25.3\% & 0.16 \\
  \midrule
  \textbf{IR: Semantic Retrieval+} & \multirow{2}{*}{28.9\%} & \multirow{2}{*}{0.17} \\
  \hspace{1em} \textbf{ \space Prefix Tree} & & \\
  \midrule
  \textbf{Few-shot LLM (Mixtral8x7B)} & & \\
  \midrule
  \hspace{1em} No context & 21.2\% & 0.13 \\
  \hspace{1em} Context & 24.0\% & 0.15 \\
  \midrule
  \textbf{Finetuned LLM (Mistral7B)} & & \\
  \midrule
  \hspace{1em} No context & 32.3\% & 0.20 \\ 
  \hspace{1em} Context & \textbf{37.0\%} & \textbf{0.23} \\ 
  \midrule
  \textbf{Upperbound} & \textbf{\upperbound} \\ 
  \bottomrule
\end{tabular}
}
\caption{\label{tab:results} QAC system results.}

\end{table}

\subsection{Prefix Trees} 

Conventional wisdom structures QAC as completing prefixes by matching the prefix to a database of known words \cite{context-qac2}, which is algorithmically solved with a trie data structure. This method constructs a tree where each node is a character that leads to other nodes which are possible continuations from that character. Traversing a trie from a root character will spell out all possible completions beginning with that character. For example, “t” leads to [“v”, “o”], and ``o'' leads to [``i'', ``a''], and so forth. Given a prefix like “to”, we follow it down the trie to ``to'' and then traverse all possible completions, which would result in complete search terms like ``toilet paper'' and ``toaster''. 

To rank the completions, we construct the trie such that each leaf node also contains the popularity of that search term, and we then take the top 10 most popular. We construct the trie on the training data's prefix-to-search-term mappings, using only cases where prefixes match the final search term.

Since the prefix tree is a memorization of training prefix/search term pairs, the theoretical success upperbound of the prefix tree on this test set is 58.9\%, which is the percent of prefix/suggestion pairs in the test set seen in the train set. We find that the basic prefix tree has a 25.3\% Success@10 and 0.16 MRR@10, reaching only 43\% of the theoretical upper bound of success for this method and only 37\% of the best QAC theoretical upperbound. 

The prefix-tree approach cannot readily incorporate context like past searches, and it cannot cover cases where the submitted search does not exactly match the typed prefix, which appears in 24.1\% of the test set. Therefore, we conclude that the QAC problem is a search term recommendation problem rather than a prefix-matching problem and requires solutions beyond basic prefix matching.

\subsection{Neural Information Retrieval} 
As neural embedding models gained popularity, various systems emerged that take advantage of embedding rather than exact word-matching for retrieval tasks (e.g., \citealt{karpukhin-etal-2020-dense,khattab2020colbert,xiong2020approximate,qu-etal-2021-rocketqa,formal2021splade}). The key benefit of semantic matching is the ability to capture related semantic intent and return search terms which do not necessarily have to start with the prefix. For example, if the prefix is “women running shoe”, a traditional system will propose only suggestions beginning with “women running shoe”. A semantic system may be able to provide alternative suggestions like “nike shoes for women” due to their semantic closeness to the prefix. This is particularly useful for the cases where no exact prefix tree match exists in the data, a scenario present in 48\% of our test cases. 

For our retriever, we use ColBERTv2 \citep{Santhanam:Khattab-etal:2022,santhanam-etal-2022-colbertv2}, a recent state-of-the-art retriever particularly suited to partial words. The ColBERT retriever first builds an index of search terms by tokenizing the terms and creating an embedding vector for each token. At inference time, the ColBERT retriever tokenizes and embeds the prefix similarly, and then computes a final score for each prefix--search term pair that takes into account similarity scores between all the token vectors in the prefix and the token vectors in the search term. In order to ensure high quality matches between prefixes of partial words and the final search terms, we append to each search term all the possible prefixes for each word in the search term. For example, if the search term is ``iphone case'', we transform it to ``iphone case i ip iph ipho iphon c ca cas'' so it contains all of its constituent prefixes. We use this semantic retriever to augment the retrieved search terms from the prefix tree when the prefix tree returns fewer than the full 10 results. We find that this semantic retrieval-augmented prefix tree outperforms the basic prefix tree matching by +3.6\% in Success@10 and +0.01 MRR.

\subsection{Off-the-shelf LLM} 

Recent QAC methods treat the problem not as information retrieval but as a generative problem, where we are tasked to generate the suggestions from a model \citep[e.g.][]{llmqac}. Recently, there has been emerging research on using LLMs in search applications in general \citep[e.g.][]{searchllm1,searchllm2}. The idea is that knowledge of what is relevant as well as the semantic relationships between prefix and search terms are accurately captured in the training of an LLM. We can then prompt the LLM and have it generate 10 relevant search terms already ranked in order. 

We first test this system using an off-the-shelf non-finetuned LLM, Mixtral-8x7B-v0.1 \citep{mixtral}. We do few-shot prompting and ask the LLM to generate a suggestion given the prefix (prompt in Appendix \ref{app:fewshot}). We perform beam search with beam size of 10 to get the top 10 suggestions from the model. We measure Success@10 and MRR@10 on the test set. We also add the past searches context in the prompt and measure the same metrics, all reported in Table~\ref{tab:results}. 

We find that few-shot prompting is able to achieve only 21.2\% success, which is worse than the basic prefix-matching system. However, including context improves the model’s performance by +2.8\%, to 24.0\%, close (but still worse) than the prefix tree. The prefix tree performs well on seen and popular prefix--search term pairs, whereas an LLM, which has no direct knowledge of past prefix/search term pairs or popularity, performs better on unseen and rarer prefix--search terms pairs -- a complete error analysis is in Appendix~\ref{app:error_analysis}. The improvement from including context is further evidence that context is important in QAC systems and suggests that LLMs can accurately capture and use the context where necessary and ignore it otherwise. 

Although the performance is slightly worse than prefix-trees, the LLM is able to incorporate context by simply inserting it into the prompt, and is able to generate a full 10 search term suggestions for 100\% of the prefixes. Overall, then, LLMs seem better suited to QAC than prefix-based approaches.

\subsection{Finetuned LLM} 

Since the previous LLM approach did not use historical prefix/search term data, the next step for generative QAC is to finetune an LLM on a zero-shot prompt using the training data, so the model can get a better understanding of the data patterns for the QAC application. The prompt we use asks the LLM to generate a suggestion given the prefix (in Appendix \ref{app:finetuned}). 

We chose Mistral-7B-v0.1 for this task \citep{mistral}. We construct the finetuning data by randomly choosing 200M prefix/search term pairs from the data and fine-tune for 10 epochs, choosing the best checkpoint by validation loss (details in the Appendix \ref{app:finetuned}). Similar to the prior approach, we decode with beam size 10 to get the top 10 suggestions in order. We also test including the context in the prompt during finetuning and testing. 

The results are reported in Table~\ref{tab:results}. We find that this setup is the best in both MRR@10 and Success@10, far surpassing the next best in success@10 by +8.1\% and MRR by +0.06. Like the off-the-shelf LLM, including context improves the model’s ability to generate the correct suggestions (+4.7\% success@10). However, we are not incorporating past notions of popularity, which means this LLM also suffers on shorter and more popular prefix/search terms. Therefore, promising avenues of exploration here involve endowing the LLM with information about prior popularity (Appendix \ref{app:error_analysis}).

\section{Conclusion}
We introduced the \datasetname\ dataset to help address a critical need for realistic, large-scale datasets for Query Autocomplete (QAC). \datasetname\ is  derived from Amazon Search logs and contains 395M examples with rich metadata. Our analysis of a range of baseline approaches suggests that QAC is a challenging context dependent task that benefits from the generative capacity of modern LLMs. In particular, finetuning LLMs to perform the QAC task and make use context leads to especially strong results. However, even the best of these systems falls well short of optimal performance on \datasetname, suggesting that there is plenty of room for further innovation.

Overall, we hope the availability of \datasetname\ helps catalyze further research and innovation in QAC, driving the development of more intuitive and efficient search functionalities across digital services.

\section{Limitations and Ethical Considerations}

In creating \datasetname, we employed a variety of methods designed to ensure user privacy, as detailed in Section~\ref{sec:prep}. We regard these steps as vital, but they do affect the data distributions in ways that are relevant. In particular, since some examples were filtered out, it is not possible to reconstruct search sessions with complete fidelity. In our experiments, we find that using search context history nonetheless leads to empirical gains, but users of the dataset should still bear in mind that it is not comprehensive as a result of ethical considerations that surround any release of naturalistic data.

\datasetname\ is derived from Amazon customer logs from the U.S.\ (Section~\ref{sec:prep}). This is a particular cultural and linguistic context that is not representative of the world population. Models and results derived from \datasetname\ should be assumed to inherent these biases. By the same token, the shopping-oriented nature of Amazon's search traffic means that \datasetname\ is unlikely to generalize to other search contexts. 

We selected our baseline models to help illuminate specific properties of the dataset and give readers a sense for the remaining headroom for system performance. Our analyses suggest that the headroom is substantial, but we recognize that different modeling choices might have led to a different assessment.

\bibliography{custom}
\clearpage

\appendix
\section*{Appendix}

\section{Data Details}

The full AmazonQAC dataset is released at \url{https://huggingface.co/datasets/amazon/AmazonQAC}.

\subsection{Training Schema}\label{app:schema}
The full schema of the training data is:
\begin{lstlisting}[numbers=none]
|query_id (string)
|session_id (string)
|prefixes (array<string>)
  |- prefix (string)
|first_prefix_typed_time (string)
|final_search_term (string)
|search_time (string)
|popularity (long)
\end{lstlisting}
The \texttt{query\_id} is a unique ID given to each row in the dataset. The \texttt{session\_id} refers to the user session ID. The \texttt{prefixes} are an array of prefix strings, in order, typed by the user to arrive at the final search term. The \texttt{first\_prefix\_typed\_time} is the timestamp of when the first prefix was typed, and the \texttt{search\_time} is the timestamp of the final search. The \texttt{popularity} is the number of times the particular search term appeared in the dataset, before filtering steps.

\subsection{Test Schema}
The full schema of the test data is:
\begin{lstlisting}[numbers=none]
|query_id (string)
|session_id (string)
|past_searches (array<array<string>>)
  |- element (array<string>)
    |- search_term (string)
    |- search_time (string)
|prefix (string)
|prefix_typed_time (string)
|final_search_term (string)
|search_time (string)
\end{lstlisting}
The \texttt{query\_id} is a unique ID given to each row in the dataset. The \texttt{session\_id} refers to the user session ID. The past searches are all searches from the session which occurred prior to the \texttt{prefix\_typed\_time}. It is an array which contains a sequence of arrays with the past search term at position 0 and the past search term's search time at position 1. The \texttt{prefix} is the current prefix string for the QAC system to take as an input. The \texttt{prefix\_typed\_time} is the time that \texttt{prefix} was typed, and the \texttt{final\_search\_term} is the final typed search term, along with the \texttt{search\_time} for that final search term.

\subsection{Regex Data Filtering}
\label{app:regex}
We apply a comprehensive regex to the data in order to filter all terms which could contain potentially sensitive personal information. For safety purposes we won't describe the details of the filters we used.

\subsection{LLM Data Filter}
\label{app:llm_filter}
After regex filtering we also apply an LLM filter step. We few-shot prompted an LLM to identify any search terms which may contain personally identifiable information or are inappropriate. Any search terms which were flagged were removed. We don't release the prompt used or LLM details for safety concerns. 

\section{Large Language Model Details}
\subsection{Few-shot LLM (Mixtral-8x7B-v0.1)}
\label{app:fewshot}

We choose Mixtral-8x7B-v0.1 as our benchmark for few-shot LLM on this task. We curate 3 examples in the prompt. For the experiment including past searches context, two of the three examples have past searches context. Our context examples are carefully chosen to show how context influences the final search term suggestion. Below is the prompt we used for no-context:

\begin{lstlisting}[language=Python,numbers=none]
### Instruction: Provide ecommerce product query suggestion starting with prefix ### Prefix: toi ### Suggestion: toilet paper ### Prefix: run ### Suggestion: running shoes for women ### Prefix: ipho ### Suggestion: iphone 15 case ### Prefix: {prefix} ### Suggestion:
\end{lstlisting}
We add context examples for the exact same search terms and prefixes in order to keep the few-shot examples consistent for both the the context and no-context experiments:
\begin{lstlisting}[language=Python,numbers=none]
### Instruction: Provide ecommerce product query suggestion related to context and starting with prefix ### Context: plunger ### Prefix: toi ### Suggestion: toilet paper ### Context: women socks, running shoes ### Prefix: run ### Suggestion: running shoes for women  ### Context: none ### Prefix: ipho ### Suggestion: iphone 15 case ### Context: {context} ### Prefix: {prefix} ### Suggestion:
\end{lstlisting}

We inference on a beam size of 10, with no sampling (0 temperature, p=1) to retrieve the top 10 generated suggestions, in order, for each prefix and context in the test set.

\subsection{Finetuned LLM (Mistral-7B-v0.1)}
\label{app:finetuned}
We chose Mistral-7B-v0.1 as our LLM to finetune. We selected a random 200M (context, prefix, final search term) triplets from the training data and placed them in the following prompt for no-context:
\begin{lstlisting}[language=Python,numbers=none]
### Instruction: Provide ecommerce product query suggestion starting with prefix ### Prefix: {prefix} ### Suggestion: {final_search_term} ###
\end{lstlisting}
We added the context for the past searches context experiment:
\begin{lstlisting}[language=Python,numbers=none]
### Instruction: Provide ecommerce product query suggestion related to context and starting with prefix ### Context: {context} ### Prefix: {prefix} ### Suggestion: {final_search_term} ###
\end{lstlisting}
We finetune the LLM using PEFT LoRA \citep{peft}, in fp16 and using the 4bit version of the model. We used a peft\_lora\_r of 256, peft\_lora\_alpha of 512, peft\_lora\_dropout of 0.05, and targeted q\_proj, k\_proj, down\_proj, v\_proj, gate\_proj, o\_proj, up\_proj, lm\_head layers. We used an AWS p3dn.24xlarge machine with 8 Tesla V100 GPUs, which took 20 hours to train 20 epochs. We cut 10M of the 200M as validation set and computed the validation loss every 500 steps, picking the best checkpoint when the validation loss stopped decreasing.

\section{Error Analysis}
\label{app:error_analysis}
We conducted an error analysis for the prefix tree and context-finetuned LLM. For the prefix tree, we found that in 16\% of cases where suggestions were provided, the correct final search term didn’t match the prefix (e.g., spelling mistakes), which the prefix tree could never get right. Generally, 74\% of cases had no match in the suggestion list due to the number of different final search terms being too large to capture in a generic top-10 popularity list which applies to all users. Other features, like personalized context, semantic matching, and prefix spelling correction, are needed to disambiguate. For the LLM, it struggles with shorter, ambiguous prefixes when no context is available likely due to not being able to use popularity information. Our analysis shows that in cases without recent context, the LLM’s Success@10 is 32\%, slightly below a basic popularity list’s 34\%. For shorter prefixes ($\leq$5 chars), the LLM performs at 13\% vs 16\% for the popularity list. Potential improvements include RAG approaches or methods to guide the LLM toward more popular suggestions.

\end{document}

%% file: table-examples.tex
\begin{table*}[tp]
    \centering
    \resizebox{1\linewidth}{!}{
    \begin{tabular}{p{1.7cm} p{1.7cm} p{3cm} p{3cm} p{3cm} p{2cm}} 
    \toprule
    Query ID & Session ID & Prefixes & First Prefix Time & Final Search Term & Search Time \\ 
    \midrule
    12 & 354 & [s, si, sin, sink, sink r, sink ra, sink rac, sink rack] & 2023-09-04T20:46:14.293Z & sink rack for bottom of sink & 2023-09-04 20:46:27 \\ 
    376 & 1886 & [a, al, alu, alum, alumi, alumi, alumin, alumin, alumind, alumind] & 2023-09-04T12:18:44.120Z & aluminum free deodorant for men & 2023-09-04 12:18:47 \\ 
    120259 & 5691 & [t, tu, tup, tupe, tupelo, tupelo ] & 2023-09-15T07:47:16.359Z & tupelo honey & 2023-09-15 07:47:20 \\ 
    983301 & 5691 & [tupelo honey, tupeo honey, tupo honey, tuo honey, to honey] & 2023-09-15T07:49:21.616Z & honey & 2023-09-15 07:49:27 \\ 
    \bottomrule
    \end{tabular}
    }
    \caption{Illustrative \datasetname\ dataset examples. The examples contain the actual prefixes that users typed on the way to selecting a search term. The session IDs and timestamps support reconstructing search contexts.}
    \label{tab:example_table}
\end{table*}

%% file: table-stats.tex
\begin{table*}[tp]
    \setlength{\tabcolsep}{10pt}
  {\begin{tabularx}{\textwidth}{l r r r r}
  \toprule
  & \multicolumn{2}{c}{\multirow{2}{*}{\textbf{Main Data}}} & \multicolumn{2}{c}{\multirow{2}{*}{\textbf{Test Data}}} \\
  \cmidrule(lr){1-1}
  \textbf{Overall} & & & & \\
  \midrule
  \hspace{1em} Data Size / \# Search Terms & 395,550,004 & - & 20,000 & - \\
  \hspace{1em} Prefixes & 4,280,432,094 & - & 20,000 & -  \\
  \hspace{1em} Unique Prefixes & 383,527,223 & 8.9\% & 15,145 & 75.7\% \\
  \hspace{1em} Unique Search Terms & 39,588,974 & 10.0\% & 16,667 & 83.3\% \\
  \hspace{1em} Unique Prefix/Search Term Pairs & 1,106,613,071 & 25.9\% & 19,871 & 99.4\% \\
  \hspace{1em} Unique Sessions & 53,839,687 & 13.6\% & 6,679 & 33.4\% \\
  \midrule
  \textbf{Patterns} & & & &  \\
  \midrule
  \hspace{1em} Average Prefix Length & 9.5 & - & 9.2 & - \\
  \hspace{1em} Average Search Term Length & 20.0 & - & 20.3 & - \\
  \hspace{1em} Average Search Term Words & 3.3 & - & 3.3 & - \\
  \hspace{1em} Search Terms Starting w/ Prefix & 344,223,609 & 87.0\% & 15,180 & 75.9\% \\
  \hspace{1em} Searches per Session & 7.3 & - & 10.3 & - \\
  \midrule
  \textbf{Train/Test Overlap} & & & &  \\
  \midrule
  \hspace{1em} Unique Prefixes Overlap & \multicolumn{2}{c}{13,375} & \multicolumn{2}{c}{88.3\%} \\
  \hspace{1em} Unique Search Term Overlap & \multicolumn{2}{c}{12,308} & \multicolumn{2}{c}{73.8\%} \\
  \hspace{1em} Unique Prefix/Search Term Overlap & \multicolumn{2}{c}{11,718} & \multicolumn{2}{c}{58.9\%} \\
  \bottomrule
\end{tabularx}}
\caption{\label{tab:analysis} Statistics on various aspects of \datasetname. We provide percentages where applicable.}
\end{table*}